\documentclass[aps,prl,twocolumn]{revtex4}
\usepackage{amssymb}
\usepackage{amsmath}
\usepackage{epsfig}
\usepackage{epstopdf}
\usepackage{color}

\begin{document}

\title{Landau-Zener Tunneling of Solitons}

\author{Vazha Loladze and Ramaz Khomeriki}
\affiliation{Physics Department, Tbilisi State University,
Chavchavadze 3, 0128 Tbilisi, Georgia}

\begin{abstract}
A simple mechanical analog describing Landau-Zener tunneling effect
is proposed using two weakly coupled chains of nonlinear oscillators
with gradually decreasing (first chain) and increasing (second
chain) masses. The model allows to investigate nonlinear
generalization of Landau-Zener tunneling effect considering soliton
propagation and tunneling between the chains. It is shown that
soliton tunneling characteristics become drastically dependent on
its amplitude in nonlinear regime. The validity of the developed
tunneling theory is justified via comparison with direct numerical
simulations on oscillator ladder system.
\end{abstract}
 \maketitle

Landau-Zener (LZ) tunneling effect \cite{landau} serves as a
powerful tool for a simple quantum-mechanical interpretation of
various fascinating wave processes in quantum and classical many
body systems. LZ model has been applied to explain transitions
between Bloch bands considering time dynamics of matter waves of
Bose-Einstein condensates in optical lattices \cite{3,4} and
acoustic waves in layered elastic structures \cite{5}. Later on the
same effect of Bloch mode transitions has been extended in spatial
domain considering optical systems with a variety of architectures:
waveguide arrays with  a step  in a refractive index \cite{6},
arrays with an applied temperature gradient \cite{7}, curved
waveguides \cite{8}, nematic crystals \cite{9}, and two-dimensional
photonic lattices \cite{10}. These macroscopic phenomena, at the
same time, has led to generalizations of original landau-Zener
problem, for instance, nonlinear LZ tunneling inducing asymmetric
transitions \cite{11,12,121}, LZ tunneling in multi-level systems
\cite{14,15} and Landau-Zener-Bloch oscillations \cite{sergej} could
be quoted among others. One can mention various application
proposals for LZ tunneling, such as targeted energy transfer
\cite{16} and all optical diode realization \cite{ram1}.

\begin{figure}[b]
\includegraphics[scale=.35]{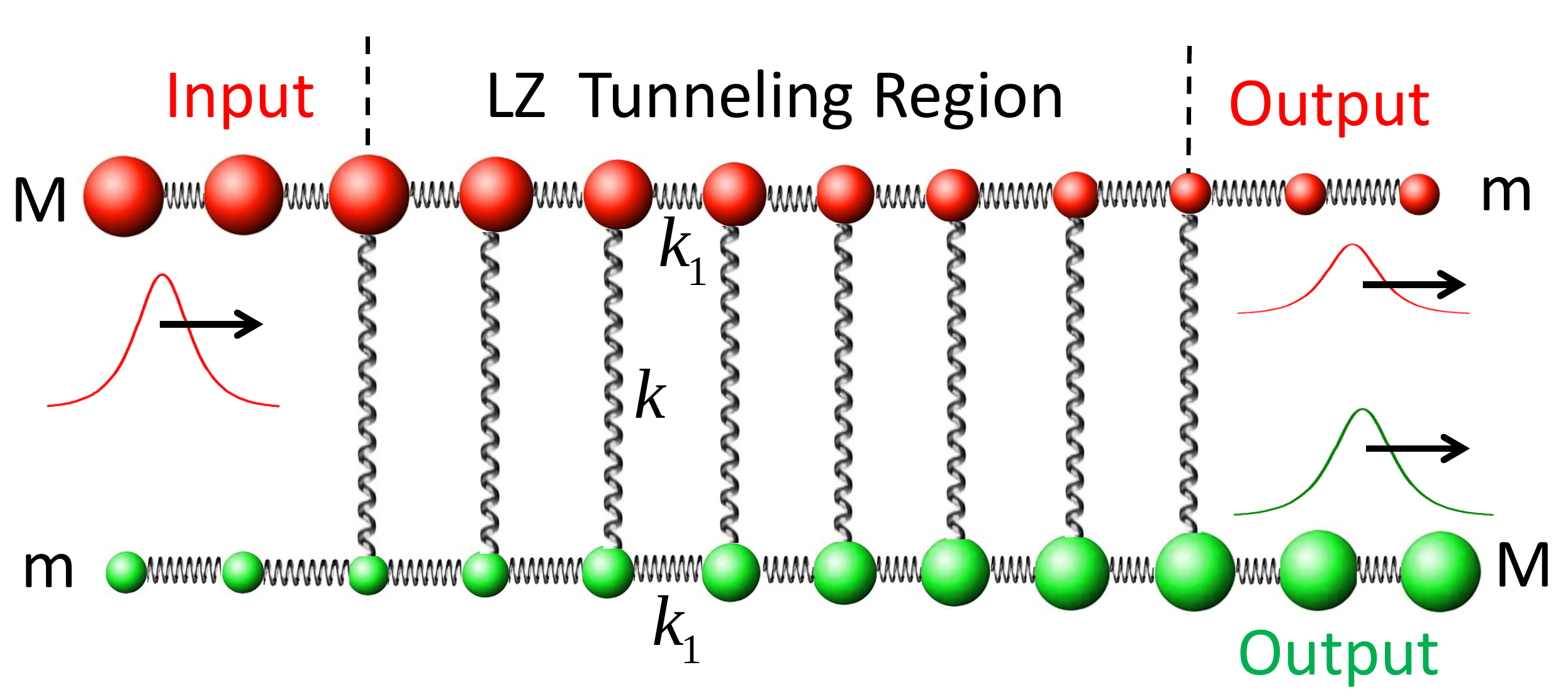}
\caption{Schematics for the oscillator ladder system. Soliton is
entering through one of the input chains and nonlinear Landau-Zener
tunneling is identified via monitoring the soliton amplitudes at the
output chains. $k$ and $k_1$ are interchain and intrachain linear
coupling constants and oscillator masses change from $M$ to $m$ in
the upper chain and vice versa in the lower chain.} \label{slide}
\end{figure}

In the present paper we consider soliton dynamics in weakly coupled
two channel system, which we interpret as LZ tunneling in
spatio-temporal domain. As an example nonlinear oscillator ladder is
examined where in tunneling region oscillator masses are varying
monotonously (decreasing and increasing along first and second
chains, respectively) as presented in Fig. \ref{slide}. This simple
conceptual scenario could be easily realized in the context of
coupled waveguides or magnetic chains with gradiented refractive
index or magnetic field, respectively. Moreover, proposed mechanism
could be applied for switching purposes between weakly coupled
completely different materials, e.g. for electrically controlled
transit of solitons from magnetic to electric parts or vice versa in
multiferroic nanostructures \cite{Jam}.

In order to support generic idea of soliton LZ tunneling we use a
most celebrated oscillator system model, namely two weakly coupled
Fermi-Pasta-Ulam chains \cite{fermi}, which consist of three parts
(Fig. \ref{slide}). Two ends of the ladder are used as input and
output and they consist of two weakly coupled FPU chains, where at
the input all oscillators have masses $M$ in upper chain and $m$ in
lower one, while, on the other hand, at the output we have masses
$m$ and $M$ in upper and lower chains, respectively. The oscillator
masses in the tunneling region depends on oscillator position via
linear law. FPU oscillator ladder with such a mass distribution
could be presented as follows:
\begin{eqnarray}
m_1(n) \ddot u_n =k_1(u_{n+1}+u_{n-1}-2u_n)+\nonumber \\
k_3(u_{n+1}-u_n)^3+k_3(u_{n-1}-u_n)^3+k(w_n-u_n), \label{111} \\
m_2(n) \ddot w_n =k_1(w_{n+1}+w_{n-1}-2w_n)+\nonumber \\
k_3(w_{n+1}-w_n)^3+k_3(w_{n-1}-w_n)^3+k(u_n-w_n) \label{112}
\end{eqnarray}
where $u_n$ and $w_n$ are displacements of n-th oscillator in upper
and lower chains, respectively. We choose mass distribution in the
tunneling region
\begin{eqnarray}
m_1(n)=m_0(1-\alpha
n), \qquad m_2(n)=m_0(1+\alpha n) \label{3}
\end{eqnarray}
such that $m_1(-N/2)=m_2(N/2)=M$ and $m_1(N/2)=m_2(-N/2)=m$ where in
tunneling region index $n$ varies in the limits $-N/2<n<N/2$; $m_0$
is an oscillator mass in the middle of ladder and $\alpha$ stands
for a mass gradient coefficient; $k_1$ is linear and $k_3$ is
nonlinear coupling stiffness of springs connecting the oscillators
of the same chain, while $k$ is a weak coupling constant between
oscillators in different chains. It should be especially mentioned
that relative difference of masses between different ends of the
same chain, i.e. the value $(M-m)/M$ should be small, otherwise
analogical to Fressnell reflection effects \cite{szameit} will take
place and one has to take into account both reflection and tunneling
processes, that makes difficult clear identification of
manifestations of LZ tunneling.

By introducing dimensionless time variable and redefining parameters
it is possible to choose $m_0=k_1=1$. Working in this setup we are
seeking the solution in the form of slow space-time modulation of
plane waves:
\begin{eqnarray}
u_n&=& \frac{A\left(\xi,\epsilon n\right)}{2}e^{i(\omega
t-pn)}+c.c., \label{333} \\
w_n&=& \frac{B\left(\xi,\epsilon n\right)}{2}e^{i(\omega
t-pn)}+c.c., \qquad \xi=\epsilon(n-vt) \nonumber
\end{eqnarray}
where $\epsilon\ll 1$ is a small expansion parameter. Collective
slow variable $\xi$ has been introduced and
$v=d\omega/dp=\sin{p}/\omega$ stands for a group velocity. Now we
suppose that $\alpha n\lesssim \epsilon$, $k \sim \epsilon$ and
$k_3\sim\epsilon$. Then in the zero approximation over $\epsilon$
substituting \eqref{333} into \eqref{111} and \eqref{112} one
automatically gets dispersion relation for plane waves
$\omega^2=2(1-\cos{p})$. While in the next approximation over
$\epsilon$ making simple phase modification for $A$ and $B$ we
obtain following equations:
\begin{eqnarray}
-i\frac{\partial{A}}{\partial{n}}=\alpha^\prime nA-\kappa
B+2r|A|^2A,
\label{51} \\
-i\frac{\partial{B}}{\partial{n}}=-\alpha^\prime nB-\kappa A+2r|B|^2
B \label{5}
\end{eqnarray}
with gradient coefficient $\alpha^\prime=\omega^2 \alpha/(2
\sin{p})$, coupling constant $\kappa=k/(2 \sin{p})$ and nonlinearity
$r=3k_3(\cos{p}-1)^2/(4 \sin{p})$. Substituting $A\sim e^{i \beta
n}$ and $B\sim e^{i \beta n}$ into \eqref{51} and \eqref{5} it is
easy to determine adiabatic levels $\beta$ for fixed $n$ and one
obtains quartic equation:
\begin{eqnarray}
(\alpha^\prime n \beta)^2=(\beta^2-\kappa^2)(r {\cal F}-\beta)^2
\label{6}
\end{eqnarray}
where ${\cal F(\xi)}=\mid A \mid^2+\mid B \mid^2$ is a conserved
quantity for fixed $\xi$.

\begin{figure}[t]
\includegraphics[scale=0.5]{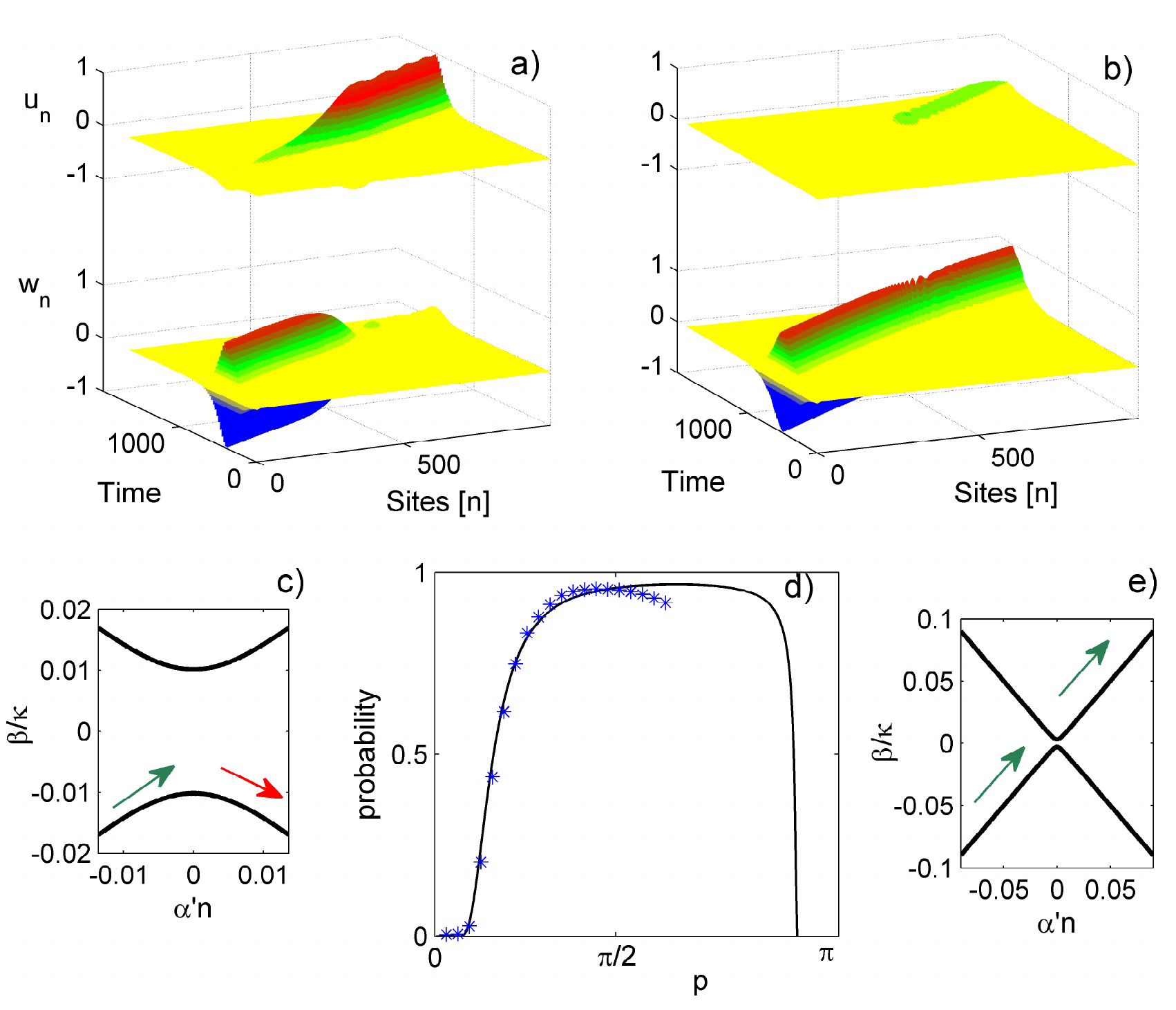}
\caption{result of numerical simulation on the initial model system
\eqref{111}, \eqref{112} in linear limit with mass distribution
\eqref{3} presented schematically in Fig. \ref{slide}. Upper surface
plots a) and b) represent linear wavepacket dynamics injected into
the lower chain. Carrier wavenumbers are $p=0.3$ and $p=\pi/2$ in a)
and b) graphs, respectively. Graph d) displays analytical dependence
of tunneling probability on carrier wavenumber given by LZ formula
\eqref{7} (solid curve) while stars are results of numerical
simulations. Graphs c) and d) represent adiabatic and diabatic
regimes in linear limit $k_3\rightarrow 0$ corresponding to the
cases in the surf plots a) and b) respectively. We use the following
parameters for the simulations and comparison: keeping intrachain
coupling constant equal to unity we choose interchain coupling as
$k=0.006$ and gradient coefficient as $\alpha=6.5\cdot 10^{-4}$.
Masses at ultimate ends of the ladder are fixed as $m=1.09$ and
$m=0.91$.}\label{zenersurf0}
\end{figure}

In case of vanishing nonlinearity $k_3\rightarrow 0$ ($r\rightarrow
0$) equations \eqref{51} and \eqref{5} reduce exactly to
Landau-Zener model \cite{landau} in the spatial domain. In the same
limit, \eqref{6} gives symmetric adiabatic levels
$\beta=\pm\sqrt{\kappa^2+(\alpha^\prime n)^2}$ displayed in Fig.
\ref{zenersurf0} c),e). According to general LZ formula
\cite{landau}, having at $n\rightarrow -\infty$ the values $A=1$ and
$B=0$, transition probability is expressed as
\begin{equation}
P=\exp\left(-\frac{\pi
\kappa^2}{\alpha^\prime}\right)=\exp{\left(-\frac{\pi k^2}{2
\omega^2 \alpha\sin{p}}\right)}. \label{7}
\end{equation}
In particular, this means that if according to \eqref{333} one has
modulated plane wave distribution at fixed $\xi=\xi_0$ and
$n=-\infty$ such that $A(\xi=\xi_0,n=-\infty)=A_0$,
$B(\xi=\xi_0,n=-\infty)=0$, then formula \eqref{6} allows to
construct the tunneling amplitudes at $n=\infty$ and the same
$\xi=\xi_0$ as follows: $|A(\xi=\xi_0,n=\infty)|^2=P|A_0|^2$ and
$|B(\xi=\xi_0,n=\infty)|^2=(1-P)|A_0|^2$.

As a result, taking initially some localized wave-function of
collective variable $\xi$, the wave will propagate through tunneling
region and at the output the amplitudes should follow to LZ
transition probability formula \eqref{6}. Particularly, we inject at
the input modulated wave via oscillating ultimate left end of the
ladder as follows $u_0(t)=\cos(\omega t)/\cosh(t/L)$, $w_0(t)=0$ or
$u_0(t)=0$, $w_0(t)=\cos(\omega t)/\cosh(t/L)$ with $L=80$ ($L$
should be large in order to have small spreading effects) and
monitor wavepacket amplitudes in both chains at the output. Fig.
\ref{zenersurf0} shows that in the range $0<p\lesssim\pi/2$
numerical experiment almost repeats theoretical curve of the
dependence of tunneling probability on the carrier wavenumber of the
injected wave-packet $p$ (see Fig. \ref{zenersurf0} d).
Particularly, the process is strongly symmetric, i.e. injecting the
wavepacket into upper (lower) chain and keeping pinned lower (upper)
chain, tunneling characteristics for both processes are exactly the
same as it should follow from original LZ model. On the other hand,
changing carrier wavenumber of the injected wavepacket from $p=0.3$
to $p=\pi/2$ one monitors transition from almost complete switch
(Fig. \ref{zenersurf0}(a) towards almost complete transmission (Fig.
\ref{zenersurf0}(b) according to general formula \eqref{7}. However,
for large wavenumbers $p\approx\pi$ the correspondence is violated
because of the reflection processes due to following reasons: for
the mentioned carrier wavenumbers the wavepacket has a small group
velocity and therefore Fressnell's reflection is in force, moreover
as one goes closer to the Brillouin zone boundary, the wavepacket
injected into the upper chain can not propagate in the same chain
due to resonance mismatch. As a result tunneling is no more
symmetric and there appear quantitative and qualitative differences
compared with the original Landau-Zener model.

\begin{figure}[t]
\includegraphics[scale=.5]{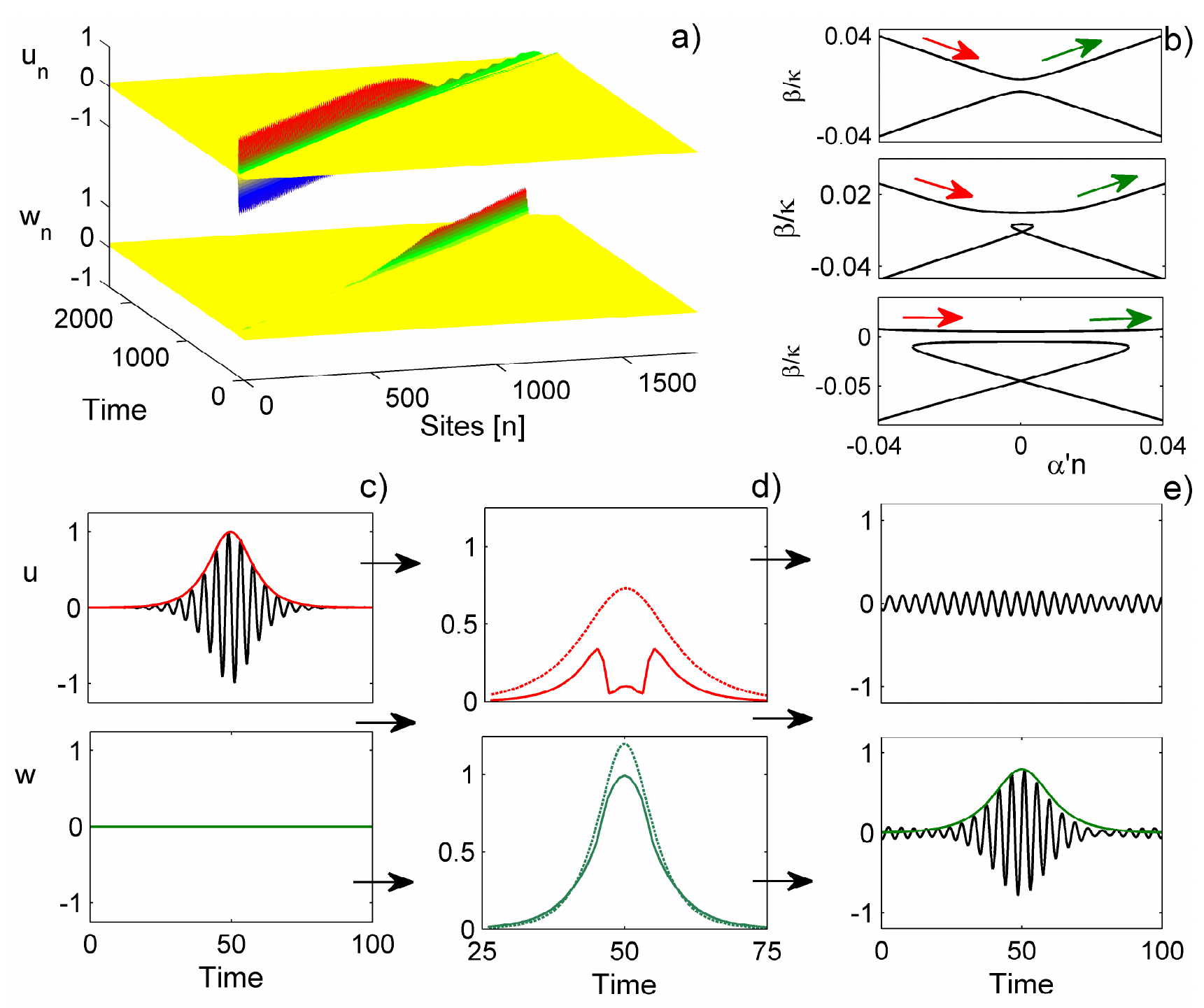}
\caption{a) surface plot of simulations on the initial model
\eqref{111}, \eqref{112} when unit amplitude soliton is injected in
the upper chain. b) displays adiabatic dynamic associated with this
process where the curves are taken solving quartic equation
\eqref{6} and arrows indicate that soliton is switching to the lower
chain. c)-d) graphs represent soliton LZ tunneling process.
Particularly, in graph c) the shape of injected soliton with
envelope function \eqref{9} in the upper chain is shown (while lower
input is pinned). In graph d) we present resulting envelopes (solid
lines) after tunneling derived from model Eqs. \eqref{51},
\eqref{5}. Dashed lines indicate constructed regular soliton
envelopes \eqref{8} with the same width at half maximum as ones
plotted by solid lines. e) shows formed output signal profiles
followed from direct numerical simulations on \eqref{111} and
\eqref{112} indicated by solid lines. Analytically computed envelope
according to scheme \eqref{10} in the lower chain is given by dashed
line, while in upper chain the envelope does not exist snce the
soliton does not form. The following parameters are used for the
calculations: nonlinearity coefficient is $k_3=0.015$, while
interchain constant takes the value $k=0.01$, gradient is
$\alpha=0.00008$ and we take carrier wavenumber $p=\pi /2$.}
\label{zenersurf1}
\end{figure}
Turning back to the nonlinear case in frames of approximate
description of Eqs. \eqref{51} and \eqref{5} we should deal with
quartic equation for $\beta$ level distribution \eqref{6}.
Corresponding curves in strongly nonlinear regime (defined by
condition $r{\cal F}>\kappa$) are displayed in Fig.
\ref{zenersurf1}b and evidently there is definite asymmetry:
Particularly, in case of small gradient constants $\alpha$ adiabatic
regime could be still realized injecting wavepacket into the upper
chain, then the system follows the upper curve of the graph b) in
Fig. \ref{zenersurf1}, while injecting the wavepacket into the lower
chain, the dynamics is always diabatic even in vanishing gradient
case $\alpha\rightarrow 0$ as it is evident from the lower curve of
the same graph. Further we will consider only such strongly
nonlinear cases $r{\cal F}>\kappa$ and examining soliton splitting
while passing through the tunneling region of the ladder.

In order to investigate soliton LZ tunneling process we employ a
weakly nonlinear soliton solution in a single oscillator chain
\begin{eqnarray}
G_n(\xi)=\frac{G\cos(\omega t-pn)}{\cosh\left(\xi\right)}, \quad
\xi=\frac{n-vt}{\Lambda},  \label{8}
\end{eqnarray}
where $G$ and $\Lambda$ are soliton amplitude and width,
respectively, and the latter is defined from the relation
$1/\Lambda=G\omega \sqrt{3k_3/2}$. Let us mentioned that the
envelope of Expression \eqref{8} is associated \cite{ramazkink} with
exact one soliton solution of nonlinear Schr\"odinger equation.

\begin{figure}[t]
\includegraphics[scale=0.5]{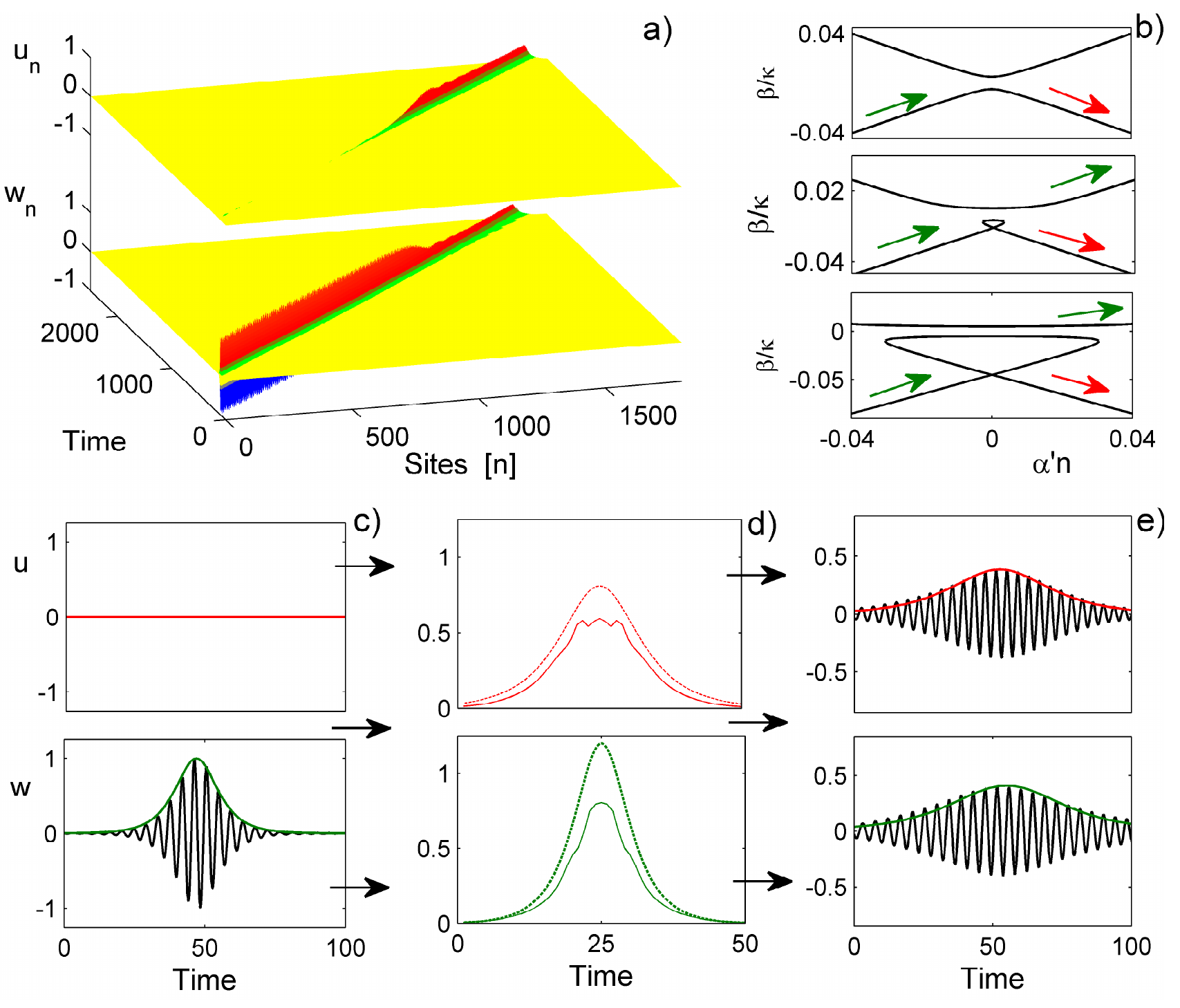}
\caption{a) Results of direct numerical simulations when unit
amplitude soliton is injected into the lower chain. b) displays
diabatic dynamic associated with the process and arrows indicate
that part of the input soliton switches to the upper chain while
another part stays in the same chain. c) the shape of injected
soliton in the lower chain is shown (while upper input is pinned).
d) represents resulting envelopes (solid lines) after tunneling
derived from model Eqs. \eqref{51}, \eqref{5}. Dashed lines indicate
constructed regular soliton envelopes \eqref{8} with the same width
at half maximum as ones plotted by solid lines. e) shows formed
output soliton profiles followed from direct numerical simulations
on \eqref{111} and \eqref{112} indicated by solid lines.
Analytically computed envelopes according to scheme \eqref{10} in
the both chains are given by dashed lines. The parameters are the
same as in Fig. \ref{zenersurf1}.}\label{zenersurf2}
\end{figure}

Now we shall demonstrate all the procedures step by step on the
particular examples presented in Figs. \ref{zenersurf1} and
\ref{zenersurf2} where injection of the soliton into upper and lower
chains, respectively, has been considered. In both cases we inject
the soliton \eqref{8} with carrier wavenumber $p=\pi/2$ (thus
carrier frequency is $\omega=\sqrt{2(1-\cos p)}=\sqrt{2}$) and we
take interchain coupling and nonlinearity constants as follows
$k=0.01$, $k_3=0.015$, while the mass gradient in the tunneling
region is $\alpha=0.00008$. First we choose the input signal with a
unit amplitude soliton \eqref{8} in the upper chain, i.e. $G^U_0=1$
and $G^L_0=0$. Corresponding surface plot and level distribution is
presented in Fig. \ref{zenersurf1}a,b, while explicit form of the
soliton shapes in upper and lower chains is presented in graph Fig.
\ref{zenersurf1}(c). This means, that according to the developed
scheme of nonlinear LZ tunneling one has following values for the
envelope variables $A$ and $B$ from \eqref{333} at the input
$n\rightarrow -\infty$:
\begin{eqnarray}
A(\xi,n\rightarrow -\infty)=\frac{1}{\cosh\left(\xi\right)}, \quad
B(\xi,n\rightarrow -\infty)=0. \label{9}
\end{eqnarray}

For each value of variable $\xi$ the input values of \eqref{9}
undergo evolution following to the nonlinear LZ equations \eqref{51}
and \eqref{5} getting after tunneling process the values
$A(\xi,n\rightarrow\infty)$ and $B(\xi,n\rightarrow\infty)$ which do
not have the regular soliton shape any more as it is evident from
graph Fig. \ref{zenersurf1}d (their shapes in both chains are
plotted as solid lines). The obtained envelope distributions
$A(\xi,n\rightarrow\infty)$ and $B(\xi,n\rightarrow\infty)$ could be
now considered as initial conditions for the associated nonlinear
Schr\"odinger equation, and the problem becomes exactly solvable
\cite{L1,L2,L3,L4}. In particular one is able to say whether the
soliton will be formed or decayed. Moreover, one can predict the
soliton amplitude and shape at the output of each chain explicitly
in a good approximation.

In this connection, first of all, one should mention that it is
crucial to determine characteristic amplitudes and widths of the
obtained distributions $A(\xi,n\rightarrow\infty)$ and
$B(\xi,n\rightarrow\infty)$. Measuring their amplitudes in Fig.
\ref{zenersurf1}d we get following values:
$G^U_1=\textrm{Max}\left[A(\xi,n\rightarrow\infty)\right]=0.995$ and
$G^L_1=\textrm{Max}\left[B(\xi,n\rightarrow\infty)\right]=0.34$,
while measuring their width at half maximum we get: $\Lambda_U=11$
and $\Lambda_L=18$. Next we should plot the regular soliton profile
\eqref{8} characterized by the same width at half maximum. For our
parameters the width of the regular soliton is defined from the
relation $1/\Lambda_0=G\sqrt{3k_3}$ and thus the amplitudes of
corresponding regular solitons are given by the following
expressions:
\begin{eqnarray}
&G^U_2=\frac{\textrm{acosh}(2)}{\Lambda_U\sqrt{3k_3}}=0.73, \quad
G^L_2=\frac{\textrm{acosh}(2)}{\Lambda_L\sqrt{3k_3}}=1.2. \label{11}
\end{eqnarray}

The latter regular solitons are displayed in both chains by dashed
lines in Fig. \ref{zenersurf1}. Comparing now the amplitudes $G_1^U$
and $G_1^L$ with $G_2^U$ and $G_2^L$, respectively, one can make
definite predictions about formation of the solitons in each chain.
In particular, as far as in the upper chain $G_1^U/G_2^U<1/2$ the
soliton will not form at the output, while in the lower chain the
soliton formation condition $G_1^L/G_2^L>1/2$ is satisfied and its
amplitude could be computed approximately as follows:
\begin{eqnarray}
G_L=2G_2^L\left(\frac{G_1^L}{G_2^L}-\frac{1}{2}\right)=0.8.
\label{10}
\end{eqnarray}
Then it is easy to recover the full shape of the solitons according
to Exp. \eqref{8} and this gives excellent fit with the results of
direct numerical simulations on initial set of equations \eqref{111}
and \eqref{112} as is evident from Fig. \ref{zenersurf1}e.

Now we proceed with the similar arguments in order to understand
soliton spitting behaviour presented in Fig. \ref{zenersurf2}a,
where unit amplitude soliton \eqref{8} is injected into the lower
chain. In this case the dynamics follows lower level line of Fig.
\ref{zenersurf2}b and therefore the process is strongly diabatic. As
a result, the picture is quite different what we have seen in case
of soliton injection into the upper chain (see Fig.
\ref{zenersurf1}). Following above developed procedure, one should
measure characteristic amplitudes of solid line curves in Fig.
\ref{zenersurf2}d. We get following values: $G^U_1=0.805$,
$G^L_1=0.6$, while for their widths at half maximum we get:
$\Lambda_U=5.6$ and $\Lambda_L=8$. Next, as in previous case, we
should plot the regular soliton profiles \eqref{8} characterized by
the same width at half maximum and similar to \eqref{11}
calculations give the reagular soliton amplitude values $G^U_2=1.18$
and $G^L_2=0.82$. Both associated regular solitons are displayed by
dashed lines in Fig. \ref{zenersurf2}. Comparing now the amplitudes
$G_1^U$ and $G_1^L$ with $G_2^U$ and $G_2^L$, respectively, one can
conclude that soliton formation condition is fulfilled both in upper
and lower chains and the solitons will form with amplitudes easily
determined from the relation \eqref{10}. Thus we get: $G^U=0.42$ and
$G^L=0.37$. Then one recovers solitons according to Exp. \eqref{8}
and compares with the results of direct numerical simulations that
is done in Fig. \ref{zenersurf2}e.

Concluding, we have identified soliton splitting phenomenon in
gradiented weakly coupled chains of nonlinear oscillators as
nonlinear Landau-Zener tunneling and made comparison between direct
numerical simulations and simple analytical scheme. The
correspondence between numerics and analytical justification becomes
worse in case of large relative mass differences and/or small
soliton propagation velocities. This is due Fresnel reflection which
has not been taken into account. The investigations of interplay
between Fresnel's reflection and Landau-Zener tunneling will be a
subject of our further studies.

\begin{acknowledgements}

The authors acknowledge financial support from Georgian SRNSF (grant
No FR/25/6-100/14). R. Kh. is supported in part by travel grants
from Georgian SRNSF and CNR, Italy (grant No 04/24) and CNRS, France
(grant No 04/01).
\end{acknowledgements}

\section{Appendix}

LZ tunneling region of our system is described by following
equation:
\begin{eqnarray}
m_0(1-\alpha n) \ddot u_n =k_1(u_{n+1}+u_{n-1}-2u_n)+\nonumber \\
k_3(u_{n+1}-u_n)^3+k_3(u_{n-1}-u_n)^3+k(w_n-u_n) \nonumber \\
m_0(1+\alpha n) \ddot w_n =k_1(w_{n+1}+w_{n-1}-2w_n)+\nonumber \\
k_3(w_{n+1}-w_n)^3+k_3(w_{n-1}-w_n)^3+k(u_n-w_n) \label{111}
\end{eqnarray}
if we redefine parameters: $k_3 \equiv \frac{k_3}{k_1}$ $k \equiv
\frac{k}{k_1}$ and introduce dimensionless time: $t \equiv t
\sqrt{\frac{k_1}{m_0}}$, we obtain:
\begin{eqnarray}
(1-\alpha n) \ddot u_n =(u_{n+1}+u_{n-1}-2u_n)+\nonumber \\
k_3(u_{n+1}-u_n)^3+k_3(u_{n-1}-u_n)^3+k(w_n-u_n) \nonumber \\
(1+\alpha n) \ddot w_n =(w_{n+1}+w_{n-1}-2w_n)+\nonumber \\
k_3(w_{n+1}-w_n)^3+k_3(w_{n-1}-w_n)^3+k(u_n-w_n) \label{112}
\end{eqnarray}
Let us seek solutions of equations \eqref{112} as follows:
\begin{eqnarray}
u_n= \frac{A(\xi,\epsilon n)}{2} e^{i(\omega t-pn)}+C.C. \nonumber\\
w_n = \frac{B(\xi,\epsilon n)}{2} e^{i(\omega t-pn)}+C.C.
\label{113} \xi=\epsilon(n-vt)
\end{eqnarray}
where $\epsilon<<1$, $v=\frac{\sin{p}}{w}$ and we suppose that
$\alpha n\lesssim \epsilon$, $k \sim \epsilon$, $k_3 \sim \epsilon$.

In the zero approximation over $\epsilon$, we have a dispersion
relation:
 \begin{eqnarray}
\omega^2=2(1-\cos{p}) \label{114}
 \end{eqnarray}
while in the linear approximation over $\epsilon$, we get:
\begin{eqnarray}
-i\frac{\partial{A}}{\partial{n}}=\alpha'nA-\kappa(B-A)+2r \mid A \mid^2 A\nonumber\\
-i\frac{\partial{B}}{\partial{n}}=-\alpha'nB-\kappa(A-B)+2r \mid B \mid^2 B\nonumber\\
\alpha'=\frac{\alpha \omega^2}{2 \sin{p}}, \kappa=\frac{k}{2
\sin{p}}, r=\frac{3}{4}(\cos{p}-1)^2k_3 \label{115}
\end{eqnarray}

with a phase transformation $A/B=A/B e^{i\kappa n}$, we arrive to
the equations:
\begin{eqnarray}
-i\frac{\partial{A}}{\partial{n}}=\alpha'nA-\kappa B+2r \mid A \mid^2 A\nonumber\\
-i\frac{\partial{B}}{\partial{n}}=-\alpha'nB-\kappa A+2r \mid B
\mid^2 B \label{116}
\end{eqnarray}
This equation coincides with the equations (5)-(6) from the main
text.

Now Let us define adiabatic levels. Substituting $A/B=A/B
e^{i(\beta+r \mathcal{F})n}$ into equations \eqref{116} (where
$\mathcal{F}=|A|^2+|B|^2$ is constant for fixed $\xi_0$), we get the
following system of equations:
 \begin{eqnarray}
\beta A=\alpha'nA-\kappa B+r (\mid A \mid^2-\mid B \mid^2) A\nonumber\\
\beta B=-\alpha'nB-\kappa A-r (\mid A \mid^2-\mid B \mid^2)B
\label{117}
 \end{eqnarray}
 from \eqref{117} we can determine:
 \begin{eqnarray}
\mid A \mid^2-\mid B \mid^2=\frac{\alpha' n \mathcal{F}}{\beta-r
\mathcal{F}} \label{118}
 \end{eqnarray}
combining  \eqref{117}-\eqref{118}, we have:
\begin{eqnarray}
(\beta-\frac{\alpha' n \beta}{\beta-r \mathcal{F}})A+\kappa B=0 \nonumber\\
\kappa A+(\beta+\frac{\alpha' n \beta}{\beta-r \mathcal{F}})B=0
\label{119}
\end{eqnarray}
equations \eqref{119} are linear homogenous equations for A and B.
We have non-trivial solutions of \eqref{119}, if:
\begin{eqnarray}
(\alpha' n \beta)^2=(\beta^2-\kappa^2)(r\mathcal{F} -\beta)^2
\label{110}
\end{eqnarray}
this quartic equations determine adiabatic levels $\beta$ for fixed
$n$.

\end{document}